\documentclass[10pt]{IEEEtran}

\usepackage{cite}
\usepackage{amsmath,amssymb,amsfonts}
\usepackage{algorithmic}
\usepackage[hidelinks]{hyperref}
\usepackage{graphicx}
\usepackage{textcomp}
\usepackage{xcolor}
\usepackage{listings}
\usepackage{subcaption}
\usepackage{authblk}
\usepackage{tabularx}
\usepackage{array}

\let\orgautoref\autoref
\renewcommand{\autoref}
{\def\sectionautorefname{Section}%
\def\subsectionautorefname{Section}%
\def\subsubsectionautorefname{Section}%
\def\figureautorefname{Fig.}%
\def\equationautorefname{Eq.}%
\orgautoref}

\usepackage{pifont}

\usepackage{xspace}
\newcommand{\etal}{\textit{et al.}~}

\newcommand{\ie}{\textit{i.e.,}~}

\newcommand{\vs}{\textit{vs.}~}
\newcommand{\one}{({\em i})\xspace}
\newcommand{\two}{({\em ii})\xspace}
\newcommand{\three}{({\em iii})\xspace}

\newcommand{\pone}{P1\xspace}
\newcommand{\atmega}{ATXMega128A1\xspace}
\newcommand{\cortex}{STM32F103REY\xspace}

\makeatletter
\renewcommand{\paragraph}[1]{\vspace*{0.03in}\noindent{\bf #1.}\hspace{0.25ex \@plus1ex \@minus.2ex}}
\makeatother

\makeatletter
\newcommand{\paragraphc}[1]{\vspace*{0.03in}\noindent{\bf #1}\hspace{1ex \@minus.2ex}}
\makeatother

\usepackage[absolute,showboxes]{textpos}

\begin{document}

\newif\ifanonymous

\title{Ageing Analysis of Embedded SRAM on a Large-Scale Testbed Using Machine Learning}

\ifanonymous
\author{Blinded for review}
\else
    \author[*, $\ddag$]{Leandro Lanzieri}
    \author[$\ddag$]{Peter Kietzmann}
    \author[$\dag$]{Goerschwin Fey}
    \author[*]{Holger Schlarb}
    \author[$\ddag$]{Thomas C. Schmidt}
    \affil[*]{Deutsches Elektronen-Synchrotron DESY, Germany $\cdot$ \{leandro.lanzieri, holger.schlarb\}@desy.de}
    \affil[$\ddag$]{Hamburg University of Applied Sciences, Germany $\cdot$ \{peter.kietzmann, t.schmidt\}@haw-hamburg.de}
    \affil[$\dag$]{Hamburg University of Technology, Germany $\cdot$ goerschwin.fey@tuhh.de}
\fi
\maketitle
\IEEEpeerreviewmaketitle

\begin{abstract}
	Ageing detection and failure prediction are essential in many Internet of Things (IoT) deployments, which operate huge quantities of embedded devices unattended in the field for years. In this paper, we present a large-scale empirical analysis of natural SRAM wear-out using 154 boards from a general-purpose testbed. Starting from SRAM initialization bias, which each node can easily collect at startup, we apply various metrics for feature extraction and experiment with common machine learning methods to predict the age of operation for this node.
Our findings indicate that even though ageing impacts are subtle, our indicators can well estimate usage times with an \(R^2\) score of 0.77 and a mean error of 24\% using regressors, and with an F1 score above 0.6 for classifiers applying a six-months resolution.
\end{abstract}


\IEEEdisplaynontitleabstractindextext


\bstctlcite{IEEEexample:BSTcontrol}

\setlength{\TPHorizModule}{\paperwidth}
\setlength{\TPVertModule}{\paperheight}
\TPMargin{5pt}
\begin{textblock}{0.8}(0.1,0.02)
     \noindent
     \footnotesize
     If you cite this paper, please use the DSD reference:
     L. Lanzieri, P. Kietzmann, G. Fey, H. Schlarb, T. C. Schmidt. Ageing Analysis of Embedded SRAM on a Large-Scale Testbed Using Machine Learning. In \emph{Proc. of DSD}, IEEE, 2023.
\end{textblock}

\section{Introduction}
An increasing multitude of embedded devices serves highly dependable duties such as driver assistance in vehicles or beam control in particle accelerators. 
To reduce cost, Commercial Off-The-Shelf (COTS) devices are often deployed, even under harsh operating conditions.
This leads to accelerated degradation and reduced lifetime of the hardware, which in turn may result in unexpected failures with critical consequences.
It is thus imperative to monitor hardware degradation \cite{lmfssw-tadmes-23} for triggering early damage minimization or scheduled maintenance.

SRAM is ubiquitously present in microcontrollers and FPGAs. 
New applications for SRAM have emerged in recent years, including random number generation~\cite{ksw-gpngi-22}, device fingerprinting \cite{hrsb-drvf-13,ksw-pcees-23}, and timekeeping \cite{rshs-tardis-12}, which leverage physical behaviours of memories.
Many of these applications rely on SRAM Physically Unclonable Functions (PUFs), which are a consequence of physical characteristics of the memory, such as cell imbalance or data retention time.
Various ageing mechanisms affect SRAM transistors, potentially changing their behaviour and impacting PUF responses or other applications relying on physical SRAM characteristics.
Monitoring hardware changes, particularly on secure and critical applications, can assist in preventing failures.
Hardware usage and health estimation via Machine Learning (ML) models can advise on preventive maintenance actions on embedded devices deployments, based on physical characteristics modifications.

In this paper, we target the question, whether SRAM ageing can be read from startup patterns. We analyse in detail the behaviour of SRAM cell initialization for a large collection of chips that naturally aged over periods from two to 18 months of board utilization time. We contribute in detail:
\begin{enumerate}
  \item a large-scale measurement study of SRAM initialization from 154 heterogeneously operated  testbed nodes
  \item a variety of subtle analyses and feature extractions from the data including blockwise averaging and spatial frequency analysis to capture age correlations
  \item  training a selection of ML models to explore their performance in estimating device usage.
\end{enumerate}

\begin{figure}[t]
    \centering
    \includegraphics[width=\linewidth]{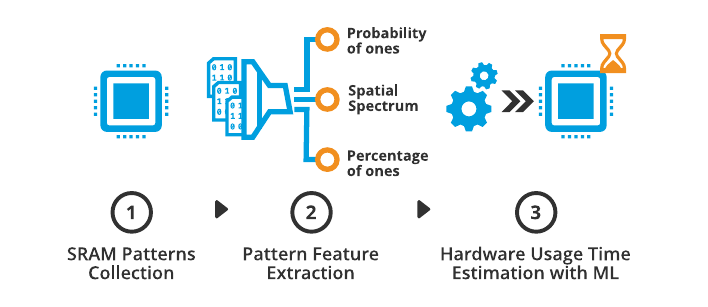}
    \caption{We gather SRAM startup patterns of embedded systems from which we derive usage-correlated features to train ML classifiers and regressors for hardware usage time estimation.}
    \label{fig:flowchart}
\end{figure}

\noindent To the best of our knowledge, this is the first work that explores \one large quantities of SRAM and \two thoroughly applies  ML techniques to estimate hardware usage time based on SRAM startup patterns, see \autoref{fig:flowchart}. Given the subtle signs of degradation in SRAM, our findings indicate good confidence in usage predictions at a resolution of 6 months. 

The remainder of this paper is structured as follows.
\autoref{sec:background} introduces the bases of the relevant hardware degradation mechanisms, the concepts of regression and classification, and discusses related work.
\autoref{sec:methodology} presents an overview of the applied methodology, while \autoref{sec:experiment} describes the analysed devices and the data collection technique.
In \autoref{sec:data_analysis}, SRAMs are analysed and features are extracted.
Subsequently, \autoref{sec:ml} evaluates the performance of various ML algorithms in estimating device usage from the selected features, while \autoref{sec:conclusion} concludes the work with an outlook.

\section{Background and Related Work}
\label{sec:background}

\subsection{SRAM Cells}
\label{subsec:sram-cells}

\begin{figure}[t]
    \centering
    \includegraphics[width=0.35\textwidth]{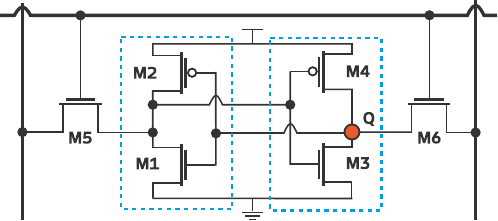}
    \caption{Diagram of a typical six-transistors SRAM cell.}
    \label{fig:6t-cell}
\end{figure}

SRAMs are used in various embedded system applications, and are preferred because they are inexpensive and easy to use.
The electronic structure of SRAMs is usually composed of 6T cells (\autoref{fig:6t-cell}) that hold one bit of data as long as power is supplied, with no need for refreshing (as opposed to DRAMs).
SRAM Cells typically consist of six Metal-Oxide-Semiconductor Field-Effect Transistors (MOSFET) forming a bistable circuit.
The circuit is formed by two cross-coupled inverters (M1, M2 and M3, M4), and has two stable states which represent 1 and 0.
Q holds the cell state, and can be accessed for reading or writing using transistors M5 and M6.

On startup, the power supply voltage ramps up, and a race condition occurs between the inverters transistors.
The MOSFET with the lowest threshold voltage activates first, setting the bit value of the cell due to the reinforcing nature of the circuit.
SRAM cells are designed to be balanced, that is, to have the same threshold voltage on all transistors.
However, due to imperfections during manufacture, differences arise in the transistors and surrounding capacitances, leading to biased cells.
As a result, most of the cells are in reality not balanced, but present a skewness towards either 1 or 0.
Given the random nature of the manufacture imperfections, the number of stable cells on a new SRAM is mostly equally split between 0 and 1.
Nonetheless, a minority of cells are unskewed, and their initial value will mainly be random and depend on thermal noise due to the similar threshold voltage of their transistors.

\subsection{Ageing Effects on SRAMs}
\label{subsec:ageing-effects}

MOSFETs undergo multiple simultaneous ageing processes during their lifetime.
Negative Bias Temperature Instability (NBTI) is one of the ageing mechanisms that affects p-type MOSFETs \cite{nbti-s,psbti-gkgr}.
NBTI is believed to occur due to the creation of traps in the oxide-substrate interface and charges in the oxide, resulting in a modification of the oxide's electrical properties.
NBTI can be modelled using the reaction-diffusion model \cite{gdrm-os}, which shows that parameters like temperature can significantly affect the impact of NBTI.
The primary consequence of NBTI is an increase of the MOSFET's threshold voltage, causing reduced switching speed of logic circuits.
MOSFETs undergo a recovery phase when the applied voltage is removed, but it is still unclear whether full recovery occurs.

Transistors in SRAM cells are affected by NBTI.
It has been reported that degradation mechanisms can reduce the static noise margin of SRAM cells by more than 15\% \cite{atac-fn}, making them more vulnerable to state flips due to noise.
Additionally, variations in the threshold voltage can directly affect the cells' data retention voltage, thus their retention time \cite{sls-qcmv-04}.
But this can be exploited to indirectly measure the device ageing, by measuring changes in the threshold voltage, effectively repurposing SRAMs as usage sensors.
As data is written into memory, transistors are stressed under different values, increasing their threshold voltage towards a direction.
For instance, when a 0 is written half of the transistors are stressed, skewing the cell to start with a value of 1.
When unskewed cells are utilized by the firmware, they start developing a bias, issuing changes on the distribution of their initial values.

\subsection{Regression and Multi-class Classification}

Supervised learning is a ML technique in which algorithms learn to predict an output based on a given input, by training on labelled data.
Algorithms are trained with a training sets, comprised of input-output pairs, and subsequently tested using a test set of samples that were not employed during training.
Supervised learning problems can be regressions or classifications.
While regressions aim is to predict a continuous value \cite{mm-rlr-20}, classifications, on the other hand, assign an output class to the provided input.
Classifications can be binary (two classes), or multi-class (more than two) \cite{a-smcm-05}.

In this work, we employ regression and multi-class classification to estimate device usage, given a group of SRAM features.
Regression models output directly the usage estimation of the devices.
Multi-class classifiers assign classes to different usage time ranges, where each class represents a range of usage time.
The goal is to predict the corresponding usage class for a given device.
Additionally, by adjusting the classes lengths we can modify the granularity of the prediction.

\subsection{Related Work}
\label{subsec:related-work}

The utilization of embedded SRAMs initial values as usage indicators has been explored in previous works.
Guo \etal proposed an approach \cite{zprs-grtf,gxrt-scar-18} to detect recycled system-on-chips (\ie used chips re-sold as new) with an enrolment process that selects cells which are likely to age.
Selected cells are later probed to test whether the device has been indeed used, based on a simple threshold comparison.
Guin \etal also detect recycled devices \cite{gwhs-drss-19}, but without an initial enrolment process.
Based on the fact that a new SRAM present a distribution of ones and zeroes centred on 50\%, they compare distribution variations with a threshold and estimate whether the chip is new.
Williams \etal enhance their recycling detection by using hardware layout information and software analysis \cite{wlph-sd-20}, which provides an insight on which cells are expected to change their balance over their lifetime.
This approach naturally requires an in-depth memory access analysis of the software, as well as a device that only executes a single binary.
These studies have focused on determining whether a device has been used or not, as this fact is enough to declare a chip as recycled.
Moreover, verification experiments have been carried out by artificially accelerated ageing.
In this work we estimate device usage beyond a binary classification, and provide insight on the variation of metrics across different usage times on devices aged on real-world deployments.

\section{Methodology Overview}
\label{sec:methodology}

Our methodology consists of three main steps as depicted in \autoref{fig:flowchart}: \one data collection (\autoref{sec:experiment}), \two SRAM analysis with feature extraction (\autoref{sec:data_analysis}), and \three evaluation of ML models for device usage estimation (\autoref{sec:ml}).
First, 1000 SRAM startup patterns from each device are collected.
By calculating the bitwise instability and probability of one, the usage-generated bias of the SRAM samples is analysed.
From these calculations, a set of metrics relevant to the device usage are identified and later used as input for ML algorithms.

Classifiers are evaluated at different prediction resolutions by discretizing the usage time into classes of \(N\) months.
\(N\) is varied between 1 and 9, which is half of the maximum usage. In effect, a 1-month resolution grants  18 classes to which a device can be assigned, while a 9-months resolution allows only two.
The hyperparameters of the models are tuned for each resolution to optimize their performance.
By applying a randomized search approach, each parameter is randomly sampled from a given distribution to obtain a set of values.
1000 combinations of parameters are created, for each of which a new model is instantiated, trained and evaluated.

To prevent overfitting the hyperparameters to the test dataset during tuning, the K-fold cross-validation technique is employed.
The training dataset is further split into K folds (\(K=5\) in our case), and each fold is used as validation set at a time, while the remaining 4 folds are used as training sets.
The process is performed K times with each fold used exactly once as the validation set.
The average performance of the model over the K folds is computed as the final evaluation metric for that specific hyperparameter combination, which is then used to select the best combination of parameters.
A final evaluation is performed on the test dataset, which is left aside for the training process, and comprises only unseen devices.

\subsection{Machine Learning Algorithms}
\label{subsec:algorithms}

\paragraphc{K-Nearest Neighbors (KNN)} algorithm estimates the output value of a query point based on the values of its nearest neighbors from the training dataset.
The distance of the training points to the query point is calculated across multiple dimensions (\ie features).
For hyperparameter tuning we varied the number of neighbors from 100 to 2000, given that in the final dataset each device contributes with 100 samples.

\paragraphc{Support Vector Machine (SVM)} algorithms attempt to find hyperplanes that separate the training data points \cite{v-nslt-99}.
For classifications, the optimal plane is the one with the largest possible margin to the points, providing the model with generalization.
This model was included in the evaluation to explore whether samples can be linearly separable in a higher-dimensional space.
The hyperparameter tuning optimized the regularization parameter \(C\) and the \(\gamma\) coefficient of the Radial Basis Function (RBF) kernel, widely employed for SVM classification problems.
The SVM classifier implementation has a one-vs-one strategy for multi-class classification, meaning that one classifier is trained for each pair of classes.

\paragraphc{Decision Trees (DTs)} predict output values by learning decision rules which are inferred from the provided features during training.
The algorithm has been included because of its interpretability, meaning that the decisions of be can be interpreted by looking at the rules.
The tuned hyperparameters for this algorithm were the maximum tree depth, the minimum amount of considered samples to add a split, and the minimum amount of features to consider on each split.

\paragraphc{Random Forest (RF)} is an ensemble algorithm that fits various DTs across data subsets \cite{b-rf-01}.
During training, the algorithm introduces randomness while creating split nodes of the trees to decrease the variance across them.
The hyperparameter space is the same as for the DTs, with the addition of the number of trees to train.

\subsection{Regression Comparison Metrics}
\label{subsec:regression_metrics}

To effectively communicate results and compare regression algorithms, uniform scores or metrics that reflect the performance of each algorithm are required.
In this work we employ the \(R^2\) score and the mean absolute percentage error (MAPE).

\(R^2\), or coefficient of determination, allows establishing a baseline model for comparison purposes \cite{w-cac-21}.
Unlike other metrics, such as the MAPE, the \(R^2\) score offers a relative metric for evaluating algorithms performance.
\(R^2\) has been used for hyperparameter tuning, and can be calculated as

\begin{equation*}
    \textit{R}^2(y,\hat{y}) = 1 - \frac{\sum_{i=1}^{n} (y_i - \hat{y_i})^2}{\sum_{i=1}^{n} (y_i - \overline{y})^2}
\end{equation*}

where \(\hat{y_i}\) is the \(i\)-th prediction, \(y_i\) is the true value, \(\overline{y}\) is the average value of the ground truth vector and \(n\) is its length.

By utilizing \(R^2\), one can determine how much better a given algorithm performs, compared to a naive model that simply outputs the mean of the data points.
The metric ranges from \(-\infty\) to 1, where \(1\) indicates that the model perfectly predicts the observed data, and a score of \(0\) indicates that the model is not better than the naive approach.
Negative values show that the model performs worse than the mean-based technique.
In summary, the higher the \(R^2\) score, the better the model's ability to explain the variation in the data from its input.

In addition, we utilize the MAPE to provide an interpretable notion of the estimation error.
The error is defined as 

\begin{equation*}
    \textit{MAPE}(y,\hat{y}) = \frac{1}{n} \sum_{i=0}^{n-1} \frac{\left\lvert y_i - \hat{y_i}\right\rvert }{\textit{max}(\epsilon, \left\lvert y_i \right\rvert)}
\end{equation*}

where \(\epsilon\) is a small number to avoid a zero-division.

\subsection{Classification Comparison Metrics}
\label{subsec:classification_metrics}

F1-score is a widely used metric that allows assessing the accuracy of classification models.
It is the harmonic mean of precision and recall, therefore it accounts for false positives and false negatives.
F1-score ranges from \(0\) to \(1\), with \(0\) being the worst possible score and \(1\) a perfect classification.
This metric is resilient against imbalanced datasets, where the distribution of samples among classes is skewed.
Although it was originally used for binary classification problems, it can be extended by calculating it for each class individually and then aggregating the results.
The aggregation is performed by calculating the metric per label and computing their average.
F1-score in a multi-class classification is described as

\begin{equation*}
    \textit{F1} = \frac{2}{N} \sum_{i=1}^{N} \frac{\textit{precision}_i \cdot \textit{recall}_i}{\textit{precision}_i + \textit{recall}_i}
\end{equation*}

\(N\) denotes the total number of classes.
\(\textit{precision}_i\) is the relation between the number of true positives and the total number of predicted positives for the \(i\)-th class, while \(\textit{recall}_i\) is the ratio between the number of true positives and the total number of actual positives.

\begin{figure}[t]
  \centering
  \includegraphics[width=\linewidth]{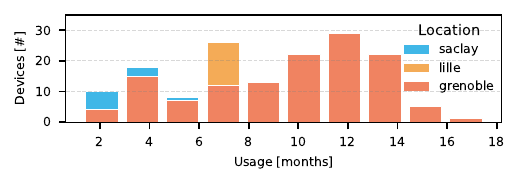}
  \caption{Usage distribution of the \cortex dataset.}
  \label{fig:m3_devices}
\end{figure}

\section{Experiment Setup}
\label{sec:experiment}

\subsection{Analysed Hardware}

The first dataset analysed in this work comprises data from  7 Microchip \atmega microcontrollers that were deployed on the Eu-XFEL at the Deutsches Elektronen-Synchrotron (DESY) in Germany \cite{xfel-06}.
The devices implement the management interface for FPGA mezzanine card carriers, which have been deployed in the accelerator as special diagnostics equipment.
These devices have been consistently running a MicroTCA management controller firmware for two years.
The sampled \atmega microcontrollers have an 8-bits AVR core with 128~KB of ROM and 8~KB of SRAM.

The second dataset analysed in this work consists of measurements from 154 embedded boards deployed on the open testbed IoT-LAB \cite{abfh-iotlab-15} in multiple locations across France.
The usage distribution of these devices is depicted in \autoref{fig:m3_devices}.
We have selected the ST \cortex microcontrollers because they are widely deployed and their SRAMs were investigated and utilized to derive PUFs for digital fingerprints in previous work \cite{ksw-pcees-23}.
These devices feature 32-bits ARM Cortex-M3 cores with 256~KB of ROM and 64~KB of SRAM.

Microcontrollers on both datasets have been operated in different ways.
Unlike the \atmega dataset, the devices deployed on the testbed execute a variety of firmwares due to the public access, therefore the memory utilization pattern is different.
This allows us to investigate the impact of hardware ageing on different usage conditions.
Most of the firmware images on the testbed microcontrollers are likely to represent typical IoT use cases, for instance, RIOT OS \cite{bghkl-rosos-18} runs a battery of tests on this testbed hardware on a weekly basis.
It is worth noting that the device usage of this dataset is calculated as the sum of the time that the devices have been booked for experiments by users.
Given that the payload of the devices is unknown, the effective usage time may differ.
This is potential a source of noise when it comes to usage estimation.

\subsection{Data Collection}
\label{subsec:data_collection}

To analyse the wear-out of SRAMs we have focused on observing the behaviour of memory bits right after the device is powered up.
Upon startup, the experiment firmware should only perform a bare initialization before transferring the SRAM values to avoid influencing bits initial states
Care should be taken to skip any initial bootstrap before transferring SRAM values, as they typically initialize memory sections.

\begin{figure}[t]
  \centering
  \includegraphics[width=0.48\textwidth]{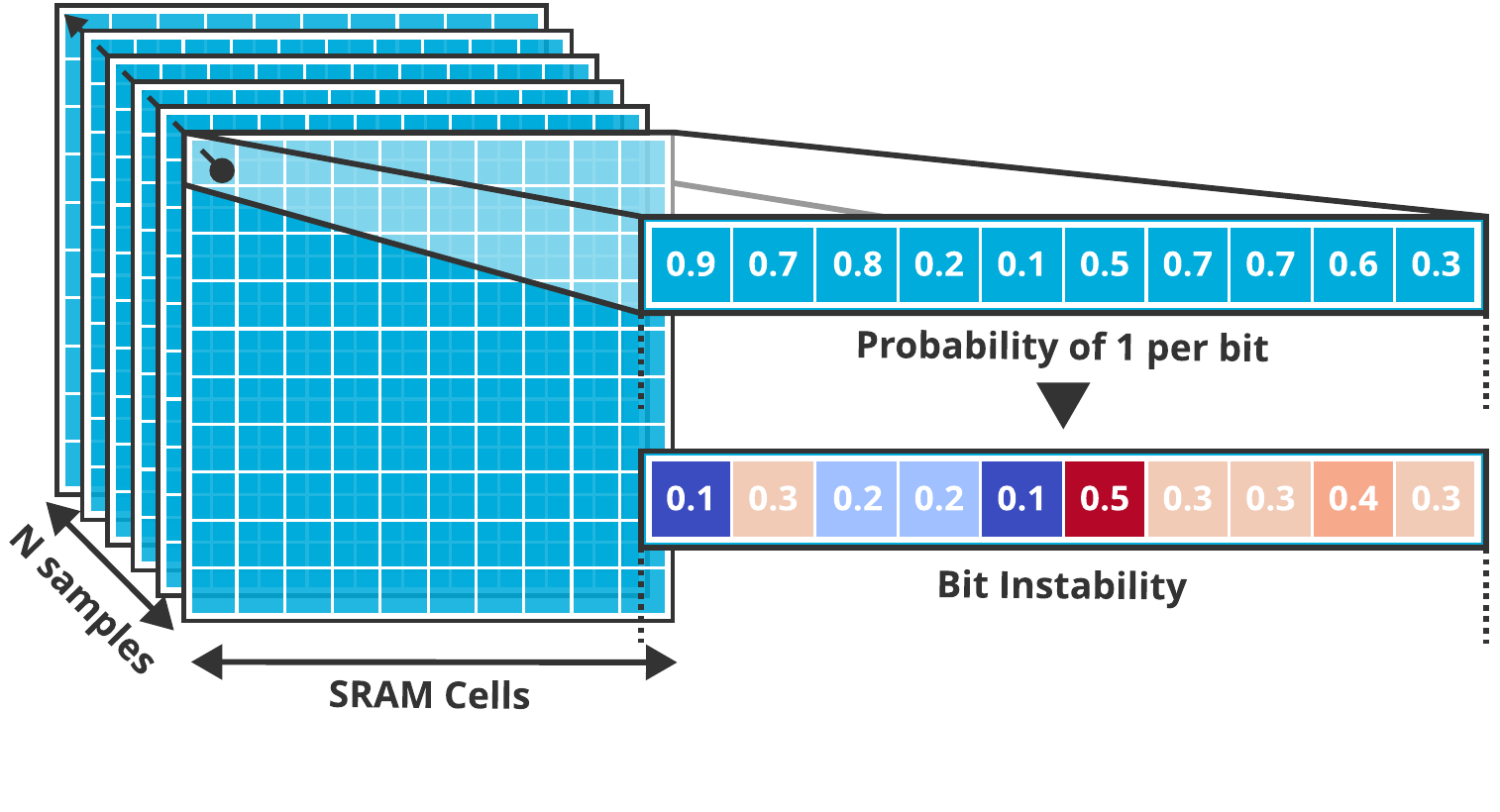}
  \caption{Probability of one and instability calculation for each bit, across multiple SRAM samples of a single device.}
  \label{fig:pone_and_activity_calculation}
\end{figure}

\begin{figure*}[t]
  \begin{subfigure}{.5\textwidth}
    \centering
    \includegraphics[width=\linewidth]{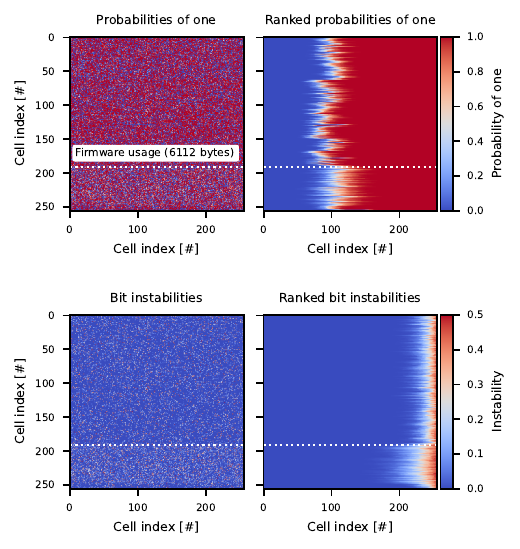}
    \caption{1005 responses of 65536 bits from an \atmega.}
    \label{fig:atxmega_instability}
  \end{subfigure}%
  \begin{subfigure}{.5\textwidth}
    \centering
    \includegraphics[width=\linewidth]{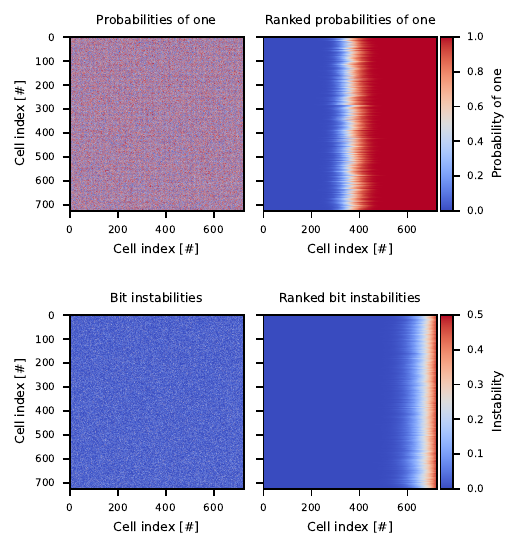}
    \caption{1000 responses of 524288 bits from an \cortex.}
    \label{fig:m3_instability}
  \end{subfigure}
  \caption{Samples from used \atmega and \cortex microcontrollers summarized using probability of ones and bit instability metrics.}
  \label{fig:devices_instability}
\end{figure*}

\subsection{SRAM Startup Patterns Datasets}

A sample in our experiment comprises a collection of ones and zeros collected from an uninitialized SRAM.
From each of the analysed devices we collect 1000 samples (both \atmega and \cortex).
Although samples from the same device are independent of one another, they share various properties and patterns.
When splitting the whole dataset into training, testing and validation, it is important to do so per-device and not per sample.
In other words, care should be taken as to avoid including samples from the same device in more than one dataset split.
Otherwise, models might overfit by learning characteristics of the device itself, preventing a generalization of the device usage.

As the device population is not uniform, datasets were stratified when splitting them.
This process attempts to keep the class frequencies (\ie device usage) approximately preserved for training, testing and validation.
This avoids overfitting models to a subset of the devices, preventing generalization.

\section{Analyzing Uninitialized SRAM Patterns}
\label{sec:data_analysis}

\begin{figure*}[t]
    \begin{subfigure}{.3\textwidth}
        \centering
        \includegraphics[width=\linewidth]{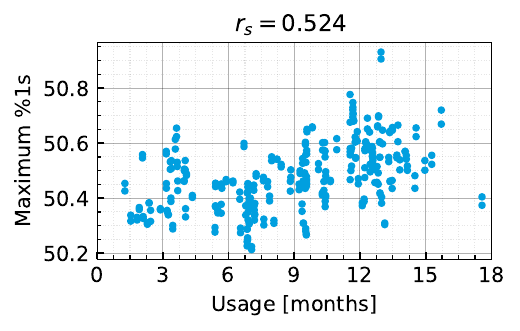}
        \caption{Maximum \%1s \vs device usage.}
        \label{fig:m3_feature_max}
    \end{subfigure}
    \hfill
    \begin{subfigure}{.3\textwidth}
        \centering
        \includegraphics[width=\linewidth]{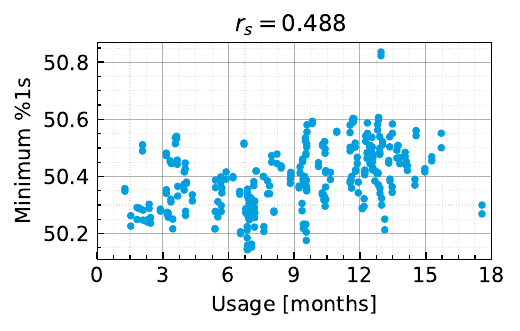}
        \caption{Mean \%1s \vs device usage.}
        \label{fig:m3_feature_mean}
    \end{subfigure}
    \hfill
    \begin{subfigure}{.3\textwidth}
        \centering
        \includegraphics[width=\linewidth]{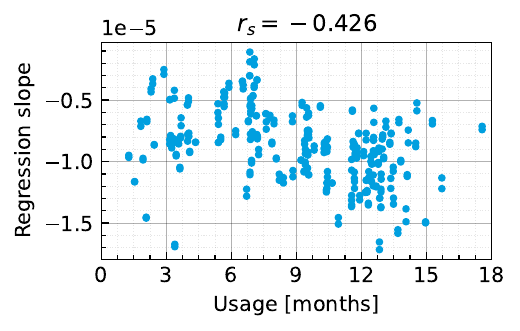}
        \caption{Minimum \%1s \vs device usage.}
        \label{fig:m3_feature_min}
    \end{subfigure}%
    \hfill
    \par\bigskip 
    \begin{subfigure}{.3\textwidth}
        \centering
        \includegraphics[width=\linewidth]{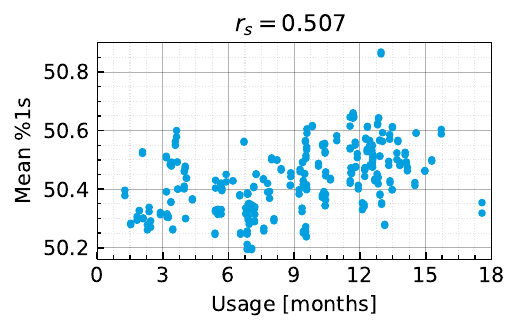}
        \caption{Intercept of the P1-address linear\\regression \vs device usage.}
        \label{fig:m3_feature_intercept}
    \end{subfigure}%
    \hfill
    \begin{subfigure}{.3\textwidth}
        \centering
        \includegraphics[width=\linewidth]{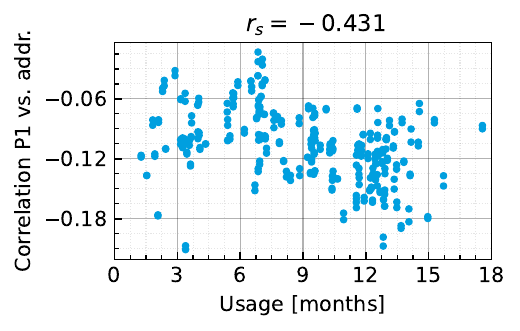}
        \caption{Slope of the P1-address linear\\regression \vs device usage.}
        \label{fig:m3_feature_slope}
    \end{subfigure}
    \hfill
    \begin{subfigure}{.3\textwidth}
        \centering
        \includegraphics[width=\linewidth]{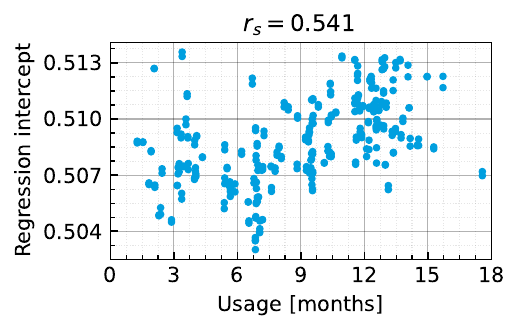}
        \caption{Spearman's correlation coefficient\\between P1 and address \vs device usage.}
        \label{fig:m3_feature_r}
    \end{subfigure}
    \caption{Selected metrics of the SRAM initial samples in relation to device usage for the \cortex microcontroller dataset, and their calculated Spearman's correlation coefficient.}
    \label{fig:m3_features_correlation}
\end{figure*}

\subsection{Cell Probability of One and Instability}
\label{subsec:cell_pone_instability}

The probability of an SRAM cell attaining the value 1 upon startup is called \emph{probability of one}, denoted \pone.
Correspondingly, the probability of a value 0 is the \emph{probability of zero} (\(P0\)), and \(\pone + P0 = 1\).
\autoref{fig:pone_and_activity_calculation} illustrates the process of obtaining the \pone of all bits of a given device.
Let $b_i(j)$ denote the bit value of cell $i$ in sampling $j$, then the \pone  of the \(i\)th bit of an SRAM ($b_i$) given N startup samples from a particular device is calculated as

\begin{equation*}
	\pone(b_i) = \frac{1}{N} \sum_{j=1}^{N} b_{i}(j)
\end{equation*}

To illustrate \pone, consider two biased SRAM cells \(b_a\) and \(b_b\), of which \(b_a\) mostly starts to 1 and \(b_b\) mostly starts to 0.
Assuming that both have a 90\% probability of initializing to their stable value, then \(\pone(b_a) = 0.9\) and \(\pone(b_b) = 0.1\).
 \pone assigns different values to both cells, even though their bias is equal.
To provide a single value that reflects how stable a cell is, we propose a symmetrized metric called \emph{bit instability} (\(I\)).
This metric ranges from 0 to 0.5 for stable and unstable cells respectively, and can be calculated from \pone as

\begin{equation*}
    I(b_i) = \begin{cases}
        \pone(b_i) & \text{if } \pone(b_i) \leq 0.5\\
        1 - \pone(b_i) & \text{if } \pone(b_i) > 0.5
    \end{cases}
\end{equation*}

\autoref{fig:devices_instability} shows the probability of ones and bit instability across 1000 samples of two SRAMs.
Plots are arranged in square matrices for ease of illustration.
Row lengths are arbitrary and have no relation with the physical memory layout on the chip.
Columns on the left of both figures present the metrics unsorted, while the right columns rank the values on each row.

Plots in \autoref{fig:atxmega_instability} show a dotted line indicating the limit of the memory usage by the firmware that the device executed continuously.
When observing the \pone for the \atmega device, it is notable that the amount of bits with a value of 0.5 is drastically reduced for the used bits.
The effect becomes even more evident when observing the bit instabilities plot.
This difference in bit instability between the used and unused sections of the memory is a clear consequence of the NBTI effect on the transistors of the SRAM cells.
Indeed, as bits are read and written during program execution, transistors are stressed towards a particular value. This increases the threshold voltage of one side of the SRAM cell and the imbalance between both sides. Correspondingly, it reduces the instability of the cell, turning the initial state of used bits  more stable.

Observing the \pone plot in \autoref{fig:atxmega_instability} also reveals that a clear majority of cells are biased towards 1 upon startup.
This indicates that more SRAM bits are stressed with a value of 0 during runtime, making it a more common value to be stored by the program.
Another reason for the higher number of cells stressed with 0 could be memory initialization. It is common practice to pre-set  variables to 0, and  C compilers initialize by default to 0 all variables of the \verb|extern| and \verb|static| storage-class, as indicated in the ANSI C standard \cite{ansi-c-11}.

\autoref{fig:m3_instability} pictures probability of ones and bit instability measurements for an \cortex microcontroller.
In this case the matrices have more than \(700 \times 700\) bits, much larger than the ATXMega memories.
Clearly the plots of this device do not show such a marked wear-out effect as the \atmega.
In particular, there is no clear boundary upon which the number of unstable cells increases.
This may be due to testbed devices executing varying firmwares while on service.
As each of the programs has a specific---and most likely different---usage of the memory, there is no single footprint imposed onto the SRAM.

\subsection{Metrics and Features with Usage Correlation}
\label{subsec:features}

In this section, we propose metrics derived from the startup SRAM patterns and analyse their correlation to device usage.
For the following analysis, the 1000 SRAM startup samples of each \cortex device are grouped into 10s to calculate their \pone.
This provides multiple samples of \pone per device, without strongly averaging out the details.
\autoref{fig:m3_features_correlation} presents the top 6 metrics that display the strongest correlation, all of which can be considered moderate \cite{m-gucc-12}.

\paragraph{Correlation Metric} To analyse and compare the selected metrics by estimating how well they are correlated to the device usage, we apply the Spearman rank-order correlation coefficient (\(r_s\)) \cite{s-pmat-87} using the SciPy \cite{scipy} implementation.
This widely used metric  measures the statistical dependence between the rankings of two variables by estimating how accurate their relationship is described by a monotonic function.
The coefficient is calculated as

\begin{equation*}
    r_s = 1 - \frac{6}{n(n^2-1)} \sum_{i=0}^{n} d_i^2 
\end{equation*}

\noindent where \(d_i\) is the difference in ranks for given values of both variables, and \(n\) is the number of samples.
As any correlation coefficient, it ranges between -1 and 1.
The absolute value indicates how strongly two variables are correlated, with 0 implying that there is no correlation.
A positive coefficient indicates that both variables increase together, while a negative one shows that a variable increases while the other decreases.
Spearman's coefficient has been chosen over Pearson's due to its robustness against extreme values and outliers \cite{m-gucc-12}.

\begin{figure}[t]
    \centering
    \includegraphics[width=\linewidth]{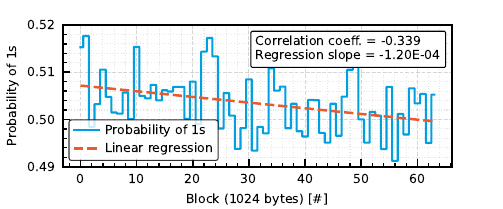}
    \caption{Probability of 1s averaged over blocks of 1024 bytes of one \cortex device SRAM.}
    \label{fig:pone_vs_address}
\end{figure}

\paragraph{Blockwise Probability of Ones} By dividing the SRAM into blocks and averaging the \pone, we can observe coarser patterns.
\autoref{fig:pone_vs_address} depicts the average \pone value for blocks of 1024 bytes of the SRAM of a \cortex device, organized by memory address.
A linear least-squares regression is fitted between both variables to highlight the data trend.
The Spearman correlation coefficient is shown to quantify the correlation between the \pone and memory addresses.
The plot reveals that blocks at the higher memory positions tend to have a lower \pone, which corresponds to a negative correlation. 
Although a wear-out of the \cortex SRAMs is expected given their usage time, the effects are not as evident at first sight with the features presented so far.

\paragraph{Percentage of Ones (\%1s) Features} This metric indicates how many of the total SRAM bits are in value 1 upon startup.
Among the features with higher correlation we find three that are related to the distribution of \%1s across multiple responses of the same device.
The maximum ((\autoref{fig:m3_feature_max}), mean (\autoref{fig:m3_feature_mean}), and minimum (\autoref{fig:m3_feature_min})) \%1s of each device display positive correlations to the usage, suggesting a slight increase in the share of ones in the responses as devices age.
This is in line with the observed behaviour on the \atmega devices---in a more subtle manner, though--- for which the used section of the memory had a stronger tendency towards ones.
This type of effect has also manifested in previous work.
Guin~\etal found in their experiments \cite{gwhs-drss-19} increments of up to 4\% in the percentage of ones after only around a week of utilization.
The marked discrepancy in the observed variations may be due to our SRAMs, which are embedded on microcontrollers and not standalone SRAM chips, so their fabrication process differed.

\paragraph{Probability of Ones (\pone) Features} Three other features in \autoref{fig:m3_features_correlation} are derived from the correlation between the \pone and the memory address.
The intercept of the linear regression between these two variables (\autoref{fig:m3_feature_intercept}) presents the highest Spearman's correlation coefficient, with a value of \(r_s=0.541\).
Additionally, \autoref{fig:m3_feature_slope} shows that the slope of the linear regression becomes more negative as devices age.
This suggests that bits in lower parts of the memory have a tendency to start with a value of 1, indicating that the payload firmware images that have been flashed on these devices tend to follow patterns with more 0s and usage in the lower addresses.
Additionally, we observe that used devices exhibit a stronger negative correlation between the \pone and the bit address than newer devices (\autoref{fig:m3_feature_r}), doubling its value in some cases.
This indicates that although subtly, cells in these areas of SRAMs age due to continuous usage.

\begin{figure}[t]
    \centering
    \includegraphics[width=\linewidth]{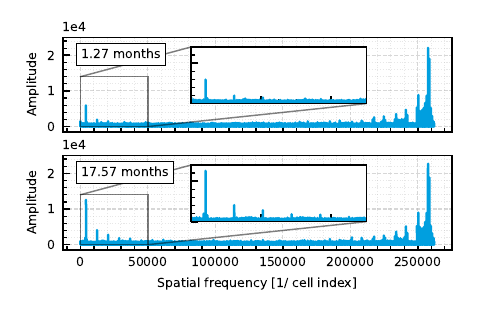}
    \caption{Probability of ones spatial spectrum of two \cortex SRAMs with different usage times.}
    \label{fig:pone_spectrum}
\end{figure}

\paragraph{Frequency Spectrum of Probability of Ones} The devices \pone in \autoref{fig:devices_instability} exhibits a noisy distribution due to the spatial randomness of process variations during cell manufacturing.
In order to reduce noise and analyse underlying structures in the data, we apply the 1-D Discrete Fourier Transform of the \pone to obtain its spatial frequency spectrum.
\autoref{fig:pone_spectrum} shows the spectra for two devices with different usage times.
We can observe equally-spaced frequencies that stand out on the extremes of the spectra and reduce their amplitude towards the middle frequencies.
Interestingly, lower frequencies show a noticeable increase on the most used device, while high-frequency amplitudes present almost no variations between spectra.
This indicates that low-frequency patterns arise with usage, while high-frequency noise remains constant.
These peaks may correspond to periodic patterns in the memory reflecting internal physical layouts of the cells on the chip.
To select frequencies of interest and reduce the number of features, we conduct a univariate feature selection.
A cross correlation between each frequency and the usage is calculated, and the 50 components with higher correlation are kept.

\section{Usage Estimation with Machine Learning}
\label{sec:ml}

\begin{figure}[t]
    \begin{subfigure}[t]{0.49\textwidth}
        \centering
        \includegraphics[width=0.95\linewidth]{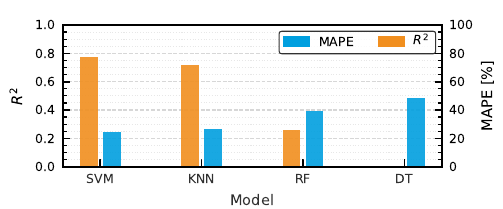}
        \caption{Regressors Mean Absolute Percentage Error and \(R^2\), evaluated with and without the spatial spectrum features.}
        \label{fig:ml_regression}
    \end{subfigure}
    \hfill
    \begin{subfigure}[t]{0.49\textwidth}
        \centering
        \includegraphics[width=0.95\linewidth]{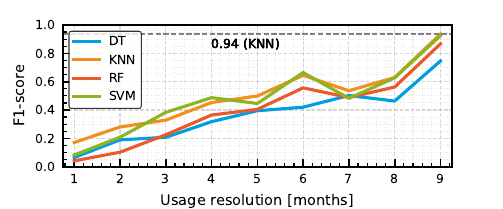}
        \caption{Multi-class classifiers F1-score for usage resolution between 1 and 9 months.}
        \label{fig:ml_classification_res_f1}
    \end{subfigure}
    \label{fig:ml_results}
    \caption{Evaluation results of regressors and classifiers.}
\end{figure}

In this section we apply different machine learning algorithms to estimate the usage time of the \cortex devices.
We evaluate the estimation performance with the common metrics presented in \autoref{subsec:regression_metrics} and \autoref{subsec:classification_metrics}.
The original dataset is split in a \(70/30\) ratio for training and testing respectively.
\autoref{tab:datasets} summarizes both dataset splits.

The inputs for all model trainings and evaluations in this section are the 56 SRAM features introduced in \autoref{subsec:features}.
It is nonetheless noteworthy that the inclusion of spatial spectrum information greatly improves model results.
Even though the frequency features present individually a lower usage correlation than the top 6, they seem to contribute together more to the end estimation scores.

\begin{table}[htbp]
    \centering
    \caption{Summary of datasets}
    \label{tab:datasets}
    {\fontsize{9}{10}\selectfont 
    \begin{tabularx}{\linewidth}{|>{\centering\arraybackslash}X|>{\centering\arraybackslash}X|>{\centering\arraybackslash}X|>{\centering\arraybackslash}X|}
        \hline
        \textbf{Dataset} & \textbf{Devices} & \textbf{Features} & \textbf{Samples} \\
        \hline
        Train & 107 & 56 & 9630 \\
        \hline
        Test & 47 & 56 & 4230 \\
        \hline
    \end{tabularx}
    } 
    \vspace{-4mm}
\end{table}

\subsection{Regression Performance}

\autoref{fig:ml_regression} presents the performance of the regressors, by showing the Mean Average Percentage Error and \(R^2\) when predicting usage times of the test dataset.
We can observe that SVM and KNN perform the best, with an \(R^2\) of \(0.77\) and \(0.71\) respectively and a percentage error around \(25\%\).
The RF regressor only yields an \(R^2\) score of \(0.25\), with a MAPE of almost \(40\%\).
Despite the random search for parameters, the DT performs worse than a naive predictor (\ie yields negative \(R^2\)) and raises the prediction error to \(48\%\).

\subsection{Classification Performance}

\autoref{fig:ml_classification_res_f1} shows the multi-class classification results
Here KNN and SVM also perform the best across most of the resolutions.
As resolutions become coarser and the number of categories decreases, F1-scores improve because models require less precision to classify a device into the correct category.
All models yield their best F1-score at a 9-months resolution, where the classification becomes binary.
KNN and SVM yield a maximum F1-score of 0.94 and 0.93 respectively, closely followed by the RF with 0.86.
As with regression, DT performs the worst for resolutions above 3 months, reaching a maximum of 0.74.
Results show that predictions with \mbox{6-months} resolutions can be achieved using SVM and KNN with a reasonable F1-score of 0.65.

\section{Conclusion and Future Work}
\label{sec:conclusion}

In this work, we studied usage effects on SRAM startup patterns of naturally-aged embedded devices.
We first analysed devices that executed a single firmware to provide insights on how used and unused sections of the SRAM are affected.
We then identified SRAM pattern features that present correlations with the device usage time, such as percentage of 1s, bitwise probability of 1s and its spatial frequency spectrum.
By training ML regressors and classifiers with the extracted features, we evaluated the application of SRAMs as inexpensive and ubiquitous device usage monitors. We found that regressions can achieve absolute errors as low as 20\% with \(R^2\) scores of 0.77, while classifiers are able to estimate usage with a 6-month resolution yielding an F1-score of 0.66.

In the future, we plan to apply deep learning techniques such as convolutional neural networks to evaluate how usage estimation can be improved.
Additionally, we aim to study how device usage time modifies other physical characteristics of embedded SRAMs such as data retention times.

\section*{Datasets Publication}
The datasets produced on work, containing the SRAM startup patterns, are publicly available under a CC Attribution 4.0 Licence with DOI:
\ifanonymous
[blinded for review].
\else
10.6084/m9.figshare.22693495.v1
\fi

\ifanonymous
\else
\section*{Acknowledgement}
We acknowledge the support by DASHH (Data Science in Hamburg - HELMHOLTZ Graduate School for the Structure of Matter) with the Grant-No. HIDSS-0002.
\fi

\bibliographystyle{IEEEtran}
\bibliography{./bib/local}

\end{document}